\input harvmac
%\draft
%-------------------------
% This paper uses harvmac
%-------------------------
\overfullrule=0pt
\def\Title#1#2{\rightline{#1}\ifx\answ\bigans\nopagenumbers\pageno0\vskip1in
\else\pageno1\vskip.8in\fi \centerline{\titlefont #2}\vskip .5in}

\def\L{{\Lambda}}
\def\p{{\partial}}

\def\CM{{\cal M}}

%\MooreFG
\lref\MooreFG{ G.~W.~Moore, ``Les Houches lectures on strings and
arithmetic,'' {\tt hep-th/0401049}.
%%CITATION = HEP-TH 0401049;%%
}
\lref\mr{R. Minasian and G. Moore, hep-th/9710230.} \lref\cl{Y.
E. Cheung and Z. Yin, hep-th/9710206.} \lref\hmg{J. Harvey, G.
Moore and M. Green, Class. Quant. Grav {\bf 14} (1997) 47,
hep-th/9605033.} \lref\cvew{C. Vafa, Nucl. Phys. {\bf B447} (1995)
252.} \lref\mms{K. Behrndt and  I. Gaida, Phys. Lett. {\bf{B401}}
(1997) 263,
 hep-th/9702168\semi  J. Maldacena, G. Moore and A. Strominger, in progress.}
\lref\suss{L. Susskind and J. Uglum, Phys. Rev. {\bf D50} (1994) 2700.}

\lref\bcovone{
M.~Bershadsky, S.~Cecotti, H.~Ooguri and C.~Vafa,
``Holomorphic anomalies in topological field theories,''
Nucl.\ Phys.\ B {\bf 405}, 279 (1993),
{\tt hep-th/9302103}.
%%CITATION = HEP-TH 9302103;%%.
}
\lref\shm{M. Shmakova, Phys. Rev. {\bf D56} (1997) 540, hep-th/9612076.}
\lref\schwarz{J. Schwarz, hep-th/9601077.}
\lref\hms{G. Horowitz, J. Maldacena and A. Strominger,
Phys. Lett. {\bf B383}, 151 (1996),
hep-th/9603109.}
\lref\jmtalk{ For a recent summary and references see J. Maldacena,
hep-th/9705078. }
\lref\afmn{I. Antoniadis, S. Ferrara, R. Minasian and K.S. Narain,
%{\it $R^4$ couplings in M and type II theories on Calabi-Yau
%spaces},
hep-th/9707013.} \lref\fks{S. Ferrara, R. Kallosh and A.
Strominger, ``${\cal N}=2$ extremal black holes'', Phys. Rev. {\bf D52},
512 (1995), {\tt hep-th/9508072}. }
%\LopesCardosoWT
\lref\dwz{
G.~Lopes Cardoso, B.~de Wit and T.~Mohaupt,
``Corrections to macroscopic supersymmetric black-hole entropy,''
Phys.\ Lett.\ B {\bf 451}, 309 (1999),
{\tt hep-th/9812082}.
%%CITATION = HEP-TH 9812082;%%
}
\lref\mohaupt{
T.~Mohaupt,
``Black hole entropy, special geometry and strings,''
Fortsch.\ Phys.\  {\bf 49}, 3 (2001),
{\tt hep-th/0007195}.
%%CITATION = HEP-TH 0007195;%%
}

%
%\IyerYS
\lref\wald{R.~M.~Wald, ``Black hole entropy in the Noether
charge,'' Phys.\ Rev.\ D {\bf 48}, 3427 (1993),
{\tt gr-qc/9307038}\semi V.~Iyer and R.~M.~Wald,
 ``Some properties of Noether charge and a proposal for dynamical black hole
entropy,'' Phys.\ Rev.\ D {\bf 50}, 846 (1994),
{\tt gr-qc/9403028}. }

\lref\dwa{ G.~Lopes Cardoso, B.~de Wit and T.~Mohaupt,
``Deviations from the area law for supersymmetric black holes,''
Fortsch.\ Phys.\  {\bf 48}, 49 (2000), {\tt hep-th/9904005}.
%%CITATION = HEP-TH 9904005;%%
}
\lref\dwb{ G.~Lopes Cardoso, B.~de Wit and T.~Mohaupt,
 ``Macroscopic entropy formulae and non-holomorphic corrections for
supersymmetric black holes,''
Nucl.\ Phys.\ B {\bf 567}, 87 (2000), {\tt hep-th/9906094}.
%%CITATION = HEP-TH 9906094;%%
} \lref\dwc{ B.~de Wit, ``Modifications of the area law and
${\cal N} = 2$
supersymmetric black holes,'' {\tt hep-th/9906095}.
%%CITATION = HEP-TH 9906095;%%
} \lref\dwd{ G.~Lopes Cardoso, B.~de Wit and T.~Mohaupt, ``Area
law corrections from state counting and supergravity,'' Class.\
Quant.\ Grav.\  {\bf 17}, 1007 (2000), {\tt hep-th/9910179}.
%%CITATION = HEP-TH 9910179;%%
} \lref\bm{K. Berndt and  T. Mohaupt,
  Phys. Rev. {\bf D56} (1997) 2206, {\tt hep-th/9611140}.}
\lref\jmcy{J. Maldacena,
%{\it N=2 extremal black holes and intersecting branes},
Phys. Lett. {\bf B403} 20, (1997),
hep-th/9611163.}
\lref\bcov{
M. Bershadsky, S. Ceccoti, H. Ooguri and C. Vafa,
``Kodaira-Spencer theory of gravity and exact results for quantum string
amplitudes,''
Commun. Math. Phys. {\bf 165}, 311 (1994), {\tt hep-th/9309140}.}

%\AntoniadisZE
\lref\naret{
I.~Antoniadis, E.~Gava, K.~S.~Narain and T.~R.~Taylor,
``Topological amplitudes in string theory,''
Nucl.\ Phys.\ B {\bf 413}, 162 (1994),
{\tt hep-th/9307158}.
%%CITATION = HEP-TH 9307158;%%
}

%\GopakumarJQ
\lref\gova{
R.~Gopakumar and C.~Vafa,
``M-theory and topological strings, I,II,''
{\tt hep-th/9809187, hep-th/9812127}.
%%CITATION = HEP-TH 9812127;%%
}

%\BreckenridgeIS
\lref\petal{
J.~C.~Breckenridge, R.~C.~Myers, A.~W.~Peet and C.~Vafa,
``D-branes and spinning black holes,''
Phys.\ Lett.\ B {\bf 391}, 93 (1997),
{\tt hep-th/9602065}.
%%CITATION = HEP-TH 9602065;%%
}

%\KatzXQ
\lref\kkv{
S.~Katz, A.~Klemm and C.~Vafa,
``M-theory, topological strings and spinning black holes,''
Adv.\ Theor.\ Math.\ Phys.\  {\bf 3}, 1445 (1999),
{\tt hep-th/9910181}\semi
T.~J.~Hollowood, A.~Iqbal and C.~Vafa,
``Matrix models, geometric engineering and elliptic genera,''
{\tt hep-th/0310272}
.}
%%CITATION = HEP-TH 9910181;%%
\lref\wittenqb{E.~Witten,
``Quantum background independence in string theory,''
{\tt hep-th/9306122}.
%%CITATION = HEP-TH 9306122;%%
}
%\NeitzkePF
\lref\nv{
A.~Neitzke and C.~Vafa,
``${\cal N} = 2$ strings and the twistorial Calabi-Yau,''
{\tt hep-th/0402128}.
%%CITATION = HEP-TH 0402128;%%
}

%\NekrasovJS
\lref\nov{
N.~Nekrasov, H.~Ooguri and C.~Vafa,
``$S$-duality and topological strings,''
{\tt hep-th/0403167}.
%%CITATION = HEP-TH 0403167;%%
}

%\SeibergRS
\lref\sw{
N.~Seiberg and E.~Witten,
``Electric-magnetic duality, monopole condensation, and confinement in
${\cal N}=2$ supersymmetric Yang-Mills theory,''
Nucl.\ Phys.\ B {\bf 426}, 19 (1994)
[Erratum-ibid.\ B {\bf 430}, 485 (1994)],
{\tt hep-th/9407087}.
%%CITATION = HEP-TH 9407087;%%
}

\lref\soka{E.~Sokatchev,
``A superspace action for
${\cal N}=2$ supergravity,''
Phys.\ Lett. \ B {\bf 100}, 466 (1981).}

%\WittenNN
\lref\wittwi{
E.~Witten,
``Perturbative gauge theory as a string theory in twistor space,''
{\tt hep-th/0312171}.
%%CITATION = HEP-TH 0312171;%%
}

\lref\klms{
 D. Kaplan, D. Lowe, J. Maldacena and  A.
Strominger, Phys. Rev.  D55  (1997) 4898,  hep-th/9609204. }
\lref\cvewoneloop{ C. Vafa and E. Witten, {\it A one loop test of
string duality}, Nucl. Phys. {\bf B447} (1995) 261,
hep-th/9505053. ???????  } \lref\kb{K. Behrndt,
G. Lopez Cardoso, B. de Wit, R. Kallosh, D. Lust and T. Mohaupt,
Nucl. Phys. {\bf B488} (1997) 236,  hep-th/9610105. } \lref\asm{A.
Strominger, ``Macroscopic Entropy of ${\cal N}=2$ black holes,''
 Phys. Lett. {\bf B383} (1996) 39,  {\tt hep-th/9602111}.  }
\lref\ascv{A. Strominger and C. Vafa,
``On the microscopic
origin of the Bekenstein-Hawking entropy,''
 Phys. Lett. {\bf B379} (1996)
99,
 {\tt hep-th/9601029}.}
\lref\berk{N. Berkovitz,
%{\it Construction of $R^4$ terms
%in N=2 D=8 superspace},
hep-th/9709116. } \lref\msw{ J.~M.~Maldacena, A.~Strominger and
E.~Witten, ``Black hole entropy in M-theory,'' JHEP {\bf 9712},
002 (1997), {\tt hep-th/9711053}.
%%CITATION = HEP-TH 9711053;%%
}
\lref\inov{A.~Iqbal,
N.~Nekrasov, A.~Okounkov and C.~Vafa,
``Quantum foam and topological strings,''
{\tt hep-th/0312022}.
%%CITATION = HEP-TH 0312022;%%
}
\lref\rov{A.~Okounkov, N.~Reshetikhin and C.~Vafa,
``Quantum Calabi-Yau and classical crystals,''
{\tt hep-th/0309208}.
%%CITATION = HEP-TH 0309208;%%
}
\lref\mms{J.~M.~Maldacena, J.~Michelson and A.~Strominger,
``Anti-de Sitter fragmentation,''
JHEP {\bf 9902}, 011 (1999),
{\tt hep-th/9812073}.
%%CITATION = HEP-TH 9812073;%%
} \lref\hlm{ G. T. Horowitz, D. A. Lowe and J. M.
Maldacena,
%{\it Statistical Entropy of Nonextremal Four-Dimensional
%Black Holes and  U-duality },
Phys. Rev. Lett. 77 (1996) 430, hep-th/9603195.} \lref\mvv{
R.~Dijkgraaf, E.~Verlinde and M.~Vonk,
``On the partition sum of the NS five-brane,''
{\tt hep-th/0205281}.
%%CITATION = HEP-TH 0205281;%%
}
\def\L{{\Lambda}}
\def\Im{{\rm Im}}
\def\Re{{\rm Re}}

\Title{\vbox{\baselineskip12pt
\hbox{hep-th/0405146}\hbox{HUTP-04/A020, CALT-68-2501}}} {\vbox{ \centerline
{Black Hole Attractors and the Topological String}}}
\bigskip

\centerline{Hirosi Ooguri,$^a$ Andrew Strominger,$^b$
and Cumrun Vafa$^b$}
\bigskip

\centerline{$^a$ California Institute of Technology, Pasadena,
CA 91125, USA}
\smallskip
\centerline{$^b$ Jefferson Physical Laboratory, Harvard University}
\centerline{Cambridge, MA 02138, USA}
\centerline{}
\bigskip
\bigskip

%\centerline{\sl Department of Physics}
%\centerline{\sl Harvard University}
%\centerline{\sl Cambridge, MA}

\bigskip
\centerline{\bf Abstract} {A simple relationship of the form
$Z_{BH}=|Z_{\rm top}|^2$ is conjectured, where $Z_{BH}$ is a
supersymmetric partition function for a four-dimensional BPS black
hole in a Calabi-Yau compactification of Type II superstring
theory and $Z_{\rm top}$ is a second-quantized topological string
partition function evaluated at the attractor point in moduli
space associated to the black hole charges. Evidence for the
conjecture in a perturbation expansion about large graviphoton
charge is given.  The microcanonical ensemble of BPS black holes
can be viewed as the Wigner function associated to the
wavefunction defined by the topological string partition
function.}

\Date{} \listtoc \writetoc
\newsec{Introduction}

Critical string theory compactified on Calabi-Yau spaces has
played a central role in both the mathematical and physical
development of modern string theory. The rich topological data, as
well as much of the low energy 4d physics associated with such
compactifications, are elegantly summarized by the $\hat c=6$
topological strings of A and B types. These topological string
theories have led to a variety of deep physical and mathematical
insights.

The physical relevance of the data provided  by the
topological string has been that it computes $F$-type terms
in the corresponding four dimensional theory \refs{\bcov, \naret}.
These
higher-derivative $F$-type terms for Type II superstring
on a Calabi-Yau are of
the general form
 \eqn\tyu{\int
d^4x d^4\theta ({ W}_{ab}{ W}^{ab})^g F_g(X^\Lambda),}
where ${ W}_{ab}$ is the graviphoton superfield of the ${\cal
N}=2$ supergravity and $X^\Lambda$ are the vector multiplet
fields.  The lowest component of ${ W}$ is $F$ the graviphoton
field strength and the highest one is the Riemann tensor. The
lowest components of $X^\Lambda$ are the complex scalars
parameterizing Calabi-Yau moduli and their highest components are
the associated $U(1)$ vector fields. These terms contribute to
multiple graviphoton-graviton scattering.  \tyu\ includes (after
$\theta $ integrations)
 an $R^2 F^{2g-2}$ term.  This fact was used in \gova , upon
considering a constant graviphoton background, to provide a
precise physical interpretation of the meaning of A-model
topological strings as computing degeneracy of BPS states
involving M2 branes wrapped around 2-cycles of Calabi-Yau upon
M-theory compactifications to five dimensions. Roughly speaking
the genus expansion of topological string encodes the spin of the
BPS state and the 2-cycle it wraps corresponds to the charge of
the M2 brane.  More precisely the topological string
partition function $Z_{\rm top}$ is the canonical ensemble
for multi-particle spinning five dimensional black holes
\petal\ as was discussed in \kkv .

In this paper we define and propose a simple and direct
relationship between the second-quantized topological string
partition function $Z_{\rm top}$ and BPS black hole partition function
$Z_{BH}$ in  four dimensions of the form\foot{Here and hereafter
the partition functions are presumed to contain $F^2(-1)^F$ type
insertions, where $F$ is an appropriate R charge, so that only BPS
states contribute.}
    \eqn\tui{Z_{BH}(p^\L,\phi^\L)=|Z_{\rm top}(X^\L)|^2,}
where $X^\L=p^\L+{i\over \pi}\phi^\L$ in a certain K\"ahler gauge.
The left hand side here is evaluated as a function of integer
magnetic charges $p^\L$ and continuous electric potentials
$\phi^\L$, which are conjugate to integer electric charges $q_\L$.
The right hand side is the holomorphic square of the partition
function for a gas of topological strings on a Calabi-Yau whose
moduli are those associated to the charges/potentials
$(p^\L,\phi^\L)$ via the attractor equations
\refs{\fks,\asm,\dwz,\dwa,\dwb,\MooreFG} . Both sides of \tui\ are
defined in a perturbation expansion in $1/Q$, where $Q$ is the
graviphoton charge carried by the black hole.\foot{The string
coupling $g_s$ is in a hypermultiplet and decouples from the
computation.} The nonperturbative completion of either side of
\tui\ might in principle be defined as the partition function of
the holographic CFT dual to the black hole, as in \ascv. We then
have the triple equality, \eqn\ttt{Z_{CFT}=Z_{BH}=|Z_{\rm
top}|^2.} This will be discussed herein for two cases:  One in
which the CFT is a 2d sigma model arising form the moduli space of
wrapped fivebranes \msw\ compactified on a circle, or the IR limit
of the $U(N)$ gauge theory on the D6 brane considered in \inov
.\foot{ This latter realization, even though it has the
disadvantage of not being conformal, it has the advantage of being
able to realize arbitrary charge states for the black hole. Having
a QFT is sufficient for computation of \ttt\ and the result (being
an index) should not depend on taking the IR limit.}

As noted already in \refs{\dwz\dwa-\dwb}, terms such as \tyu\ are
nonvanishing in the presence of a black hole and therefore correct
the BPS entropy. It is plausible that, as discussed in
\refs{\dwz\dwa-\dwb}, due to supersymmetric nonrenormalization
theorems, these $F$-terms are the only ones which correct the
(indexed) BPS entropy. We derive \tui\ building on the
supergravity analysis of \refs{\dwz \dwa-\dwb}. A key new point is
that the supergravity partition function naturally produces a
mixed thermodynamic ensemble, in which magnetic charges $p^\L$ are
fixed integers, while electric charges $q_\L$ are summed over
weighted by the electric potential as $e^{-\phi^\L q_\L}$. This
distinction is irrelevant to the computation of the leading order
area-entropy relation. However it plays a crucial role in
disentangling higher order corrections and achieving the simple
relation \tui.

The existence of fundamental connection between four-dimensional
BPS black holes and the topological string might have been
anticipated from the following observation. Calabi-Yau spaces have
two types of moduli: K\"ahler and complex structure. The
worldsheet twisting which produces the A (B) model topological
string from the critical superstring eliminates all dependence on
the complex structure (K\"ahler) moduli at the perturbative level.
Hence the perturbative topological string depends on only half the
moduli. Black hole entropy on the other hand, insofar as it is an
intrinsic property of the black hole, cannot depend on any
externally specified moduli (this statement is corrected by the anomaly
 as we
discuss in the next paragraph). What happens at leading order is that the moduli
in vector multiplets are driven to attractor values at the horizon
which depend only on the black hole charges and not on their
asymptotically specified values. Hypermultiplet vevs on the other
hand are not fixed by an attractor mechanism but simply drop out
of the entropy formula. Consistency suggests this mechanism should
extend beyond leading order, and some evidence for this has been
given in \refs{\dwz,\dwa,\dwb}.  It is natural to assume this is valid
to all orders in a $1/Q$ expansion.
Hence the perturbative topological string and
the large black hole partition
functions depend on only half the Calabi-Yau moduli. It would be
surprising if string theory produced two functions on the same
space that were not simply related. Indeed we shall argue that
they are simply related as in \tui.

The fact that the {\it mixed ensemble},
rather than microcanonical ensemble, should correspond to topological
string is very natural. The very formulation of mixed ensemble
depends on how one splits the charges to electric/magnetic pairs.
Indeed topological string partition function also
depends on the choice of how one splits electric and magnetic charges,
as was discussed in \wittenqb .  Choice of electric versus
magnetic is like a choice of position versus momentum
for topological strings.  In fact purely from this perspective
and the commutation relations between electric and magnetic
charges and the corresponding chemical potentials one can derive
the fact that the imaginary part of the attractor moduli should
be identified with the chemical potentials.  Furthermore according
to \wittenqb\ $Z_{\rm top}$ behaves
as a wave function $Z_{\rm  top}\sim
\psi (x)$ in a given basis and going from electric
to magnetic leads to its Fourier transform
$\tilde \psi (p)$.   In this context
the formula we are proposing $Z_{BH}=|Z_{\rm top}|^2
=|\psi|^2$ becomes even more
suggestive and intriguing:  The probability density of the topological
wave function $\psi$ is being related to the number density of black
holes of a fixed magenetic charge and fixed electric chemical potential.
More precisely we can view the black hole microcanonical ensemble
as the {\it Wigner function} on the electric/magnetic charge space
determined by the topological string wave function.  This is one
of the most intriguing consequences of the present paper.
Aspects of this relation is discussed in section 6.

In the preceding discussion and most of the paper we ignore
several important subtleties.  It is known that the degeneracy of
actual single particle states can jump as we change the vevs for
the scalars in the vector multiplets \sw . One might try to avoid
this by dropping the restriction to single black hole states in
$Z_{BH}$, as we shall ultimately find appropriate. However this
does not avoid the problem because the asymptotic values of the
moduli still enter the index due to the noncompact center-of-mass
integrations. Thus even the perturbative entropy would not be
expected to depend just on the charges of the black hole, but also
the background the black hole is embedded in. In fact this is
consistent with another fact for topological strings which
reinforces \tui :  the holomorphic anomaly!  It is known that the
topological string partition function $Z_{\rm top}$ {\it does} depend
on the background $(t_B,t_B^*)$ we expand about, and not just on
the chiral vector multiplet $X$ \refs{\bcovone,\bcov}.
In \tui\ we identify, via
the attractor mechanism, the $X$ with magnetic charge and
electric potential. If we identify $(t_B,t_B^*)$ with the
background where the black hole entropy is computed in, then
 \tui\ is consistent
with the fact that {\it both} sides do depend on the choice of the
background.  We thus expect that both sides of \tui\ to depend in
addition on vector and possibly hypermultiplets of the background (the
moduli at infinity) where the black hole is embedded in. Finally
we note that the holographic dual partition function $Z_{CFT}$ is
also likely to acquire background dependence via non-compact
Coulomb branches. A partial and inconclusive discussion of these
issues appear in section 7. A particualrly interesting issue is
whether or not the partition functions depend on hypermultiplet
vevs at the nonperturbative level via D-brane instantons as in
\refs{\wittwi,\nv ,\nov}.

To summarize, we have suppressed some subtle and important issues
which are ultimately essential for a complete understanding of the
meaning of \tui. Our hope is that the naive analysis presented
here has suggested the correct exact relation, in which both sides
of \tui\ acquire background dependence. We leave this to future
investigations.

The relation \tui\ states that the supersymmetric partition
function of BPS black hole microstates equals the squared
partition function of a second-quantized gas of topological
strings on a Calabi-Yau. Our derivation is mechanical in nature
and relies on a detailed comparison of the perturbation expansion
of both quantities.  (We propose a streamlined derivation of this
perturbative statement in section 3.3.) However we feel that such
a simple physical relationship should have an equally simple
physical derivation which extends beyond perturbation theory.

This paper is organized as follows:  In section 2 we discuss the
supergravity aspects of F-term corrections.   There we motivate
the relation $Z_{BH}=|Z_{\rm top}|^2$ and the interpretation of
$Z_{BH}$ as a mixed ensemble.  In section 3 we discuss some
normalization issues for relating this to topological string
amplitudes and propose an alternative simple, but  non-rigorous
derivation of the main formula $Z_{BH}=|Z_{\rm top}|^2$. Section 4
discusses two choices for dual CFT's. In Section 5 we discuss the
relation between our proposal with the proposed relationship
between topological string amplitudes and crystal melting. Section
6 comments on the interpretation of $Z_{\rm top}$ as a quantum
amplitude. In section 7 we discuss how the entropy of black hole
as well as the topological string amplitudes do depend in addition
on vector multiplet background (through holomorphic anomaly) and
possibly hypermultiplet background (through non-perturbative
corrections).

\newsec{Supergravity}
\subsec{The ${\cal N}=2, d=4$ area-entropy formula} In this subsection, we
review the leading semiclassical area-entropy formula for a
general ${\cal N}=2,~d=4$ extremal black hole characterized by magnetic
and electric charges $(p^\L,q_\L)$ derived in \refs{\fks,\asm} and
recently reviewed in \MooreFG. The asymptotic values of the moduli
in vector multiplets, parameterized by complex projective
coordinates $X^\L, ~\L=0,1,\dots,n_V$, in the black hole solution
are arbitrary. These moduli couple to the electromagnetic fields
and accordingly vary as a function of the radius. At the horizon
they approach an attractor point whose location in the moduli
space depends only on the charges. The locations of these
attractor points can be found by looking for supersymmetric
solutions with constant moduli. They are determined by the
attractor equations,\foot{ As in \msw, we have redefined $C, ~X$
and $F$ by factors of $e^{K/2}$ relative to \asm\ so that
$(X^\L,F_\L)$ is a holomorphic section. At the attractor point our
variables are related to those of \dwa\ by
$CX^L_{us}=-2iY^\L_{them},~~~~CF_{\L}^{us}(X^\L)=-2iF_{\L}^{them}(Y^\L),~~~~C=-2i\bar
Z e^{K/2}$.} \eqn\prf{p^\L=\Re[CX^\L],}
\eqn\qrf{q_\L=\Re[CF_{0\L}],} where $F_{0\L}=\p F_0/\p X^\L$ are the
holomorphic periods, and the subscript $0$ distinguishes these
from the string loop
corrected periods to appear in the next subsection.
%\foot{
%For IIA string theory compactified on a Calabi-Yau threefold, the
%leading prepotential in our conventions is
%\eqn\fcy{F_0(X)=D_{ABC}{X^AX^BX^C \over X^0}} where $A,B=1,..n_V$.
%The intersection numbers are \eqn\dabc{6D_{ABC}\equiv \int_M
%\a_A\wedge\a_B\wedge\a_C ,} where the $\a_A$ are an integral basis
%for $H^2(M;{\bf Z})$.}
Both $(p^\L,q_\L)$ and $(X^\L,F_{0\L})$
transform as vectors under the $Sp(2n+2;Z)$ duality group.

The $(2n_v+2)$ real equations \prf\ and \qrf\ determine the
$(n_v+2)$
complex quantities $(C,X^\L)$ up to K\"ahler transformations, which
act as \eqn\oopf{ K\to K-f(X)-\bar f(\bar X),~~~X^\L \to
e^fX^\L,~~~ F_0 \to e^{2f}F_0,~~~ C \to e^{-f}C,} where the K\"ahler
potential $K$ is given by \eqn\cgg{e^{-K}=i\bigl(\bar X^\L
F_{0\L}- X^\L \bar F_{0\L}\bigr).}  We could at this point set
$C=1$ and fix the K\"ahler gauge but later we shall find other
gauges useful. It is easy to see that - as required - the charges
$(p^\L,q_\L)$ determined by the attractor equations \prf, \qrf\
are invariant under K\"ahler transformations. Given the horizon
attractor values of the moduli determined by \prf\ and \qrf\ the
Bekenstein-Hawking entropy may be written \eqn\ent{S_{BH}={1 \over
4}{\rm Area} ={\pi } |Q|^2 ,} where $Q=Q_m + i Q_e$ is a
complex combination of the magnetic and electric graviphoton charges
and \eqn\ftyv{|Q|^2={i \over 2} \bigl( q_\L \bar C \bar
X^\L-p^\L \bar C \bar F_{0\L}\bigr)={C\bar C \over 4} e^{-K}.} The
normalization of $Q$ here is chosen so that $|Q|$ equals the radius
of the two sphere at the horizon.

It is useful to rephrase the above results in the context of type
IIB superstrings in terms of geometry of Calabi-Yau.  In this case
the attractor equations fix the complex geometry of the
Calabi-Yau.  The electric/magentic charges correlate with three
cycles of Calabi-Yau.  Choosing a symplectic basis for the three
cycles gives a choice of the splitting to electric and magnetic
charges.  Let $A_\L$ denote a basis for the electric three cycles,
$B^\Sigma $ the dual basis for the magnetic charges and $\Omega$
the holomorphic 3-form at the attractor point. $\Omega$ is fixed
up to an overall multiplication by a complex number $\Omega
\rightarrow \lambda \Omega$. There is a unique choice of $\lambda$
such that the resulting $\Omega$ has the property that
\eqn\ola{p^\L=\int_{A_\L}{\rm Re}\Omega={\rm Re} [CX^\L] \qquad
q_\L=\int_{B^\L }{\rm Re}\Omega={\rm Re}[CF_{0\L}]}
where ${\rm Re} \Omega={1\over 2} (\Omega+{\overline \Omega})$.
In terms of this choice the black hole entropy can be written as
\eqn\dss{S_{BH}={\pi \over 4} \int_{CY} \Omega \wedge {\overline
\Omega}}

\subsec{Higher order corrections} In this subsection we review
some results of \refs{\dwz, \dwa, \dwb, \dwd}, in which higher
derivative corrections, suppressed at large charges, to the
entropy\foot{The term entropy in \refs{\dwz, \dwa, \dwb, \dwd} and
here more precisely should refer to an indexed quantity protected
from some types of corrections by supersymmetry. This is discussed
in section 4.3 below.} formula \ent\ are derived from the $F$-term
corrections to the supergravity action using Wald's method \wald.
(For a comprehensive review of the results of these papers and
related subjects, see \mohaupt.) All other corrections to the
action involve hypermultiplets. Hypermultiplet vevs are not fixed
at the horizon and can be varied continuously. If the entropy
depended on hypermultiplet vevs, it could not be an intrinsic
property of the black hole and the number of microstates for fixed
charges could not be an integer. Hence we expect the vector
multiplet $F$-terms to encode all corrections to the entropy.

$F$-term corrections to the action are encoded in a string loop
corrected holomorphic prepotential \eqn\dxvl{F(X^\L,W^2)=
\sum_{h=0}^\infty F_h(X^\L)W^{2h},} where $F_h$ can be computed by
topological string amplitudes (as we review in the next section)
and $W^2$ involves the square of the anti-self dual graviphoton
field strength. This obeys the homogeneity equation
\eqn\efx{X^\L\p_\L F(X^\L,W^2)+W\p_{W} F(X^\L,W^2)=2 F(X^\L,W^2).}
Near the black hole horizon, the attractor value of $W^2$ obeys
\eqn\fjy{C^2W^2=256.} The exact attractor equations are then
simply \eqn\psrf{p^\L=\Re[CX^\L],}
\eqn\qrsf{q_\L=\Re\left[CF_\L\left(X^\L,{256 \over C^{2}}\right)
\right].} This is essentially the only possibility consistent with
symplectic invariance. It is then argued in \refs{\dwz, \dwa,
\dwb, \dwd}\ that the entropy as a function of the charges is
\eqn\fdc{S_{BH}={\pi i \over 2}(q_{\L}\bar C\bar X^\L-p^\L \bar
C\bar F_{\L})+{\pi \over 2}{\Im}[C^3\p_C F] ,} where $F_\L,~X^\L$
and $C$ are expressed in terms of the charges using \psrf\ and
\qrsf.

 \subsec{Reinterpretation of
the corrected entropy formula}

In this subsection we reinterpret the $F$-term corrected entropy
formula \fdc, and thereby cast it in a significantly simpler form.

Let us introduce $\phi^\Lambda$ as the imaginary parts of
$CX^\Lambda$ and write
\eqn\chemical{
  C X^\Lambda = p^\Lambda + {i \over \pi} \phi^\Lambda ,}
where we applied the first half of the attractor equations \psrf\
for the real parts of $CX^\Lambda$. Defining a
function ${\cal F}$ of $(\phi, p)$ by
\eqn\calf{
\eqalign{ {\cal F}(\phi , p ) &
= -\pi\ {\rm Im}\left[ C^2 F\left(X^\Lambda, {256 \over C^2}\right)
\right] \cr
& =- \pi {\rm Im} \left[ F\left( CX^\Lambda, 256\right)\right]\cr
& = - \pi {\rm Im} \left[
F\left( p^\Lambda + {i \over \pi} \phi^\Lambda, 256\right) \right]
,}}
the second half of the attractor equations \qrsf\
can be written as
\eqn\jji{q_\L=\half(CF_\L+ \bar C\bar
F_\L) = -{\partial \over \partial \phi^\L}
{\cal F}(\phi , p),}
where \eqn\hyy{{\p \over \p \phi^\L}={i\over \pi C}{\p \over \p
X^\L}-{i \over \pi \bar C}{\p \over \p \bar X^\L}.}

Using the homogeneity relation,
\eqn\lpo{C\p_C F\left(X^\Lambda, {256 \over C^2}\right)
=X^\L {\partial \over \partial X^\L} F -2F,}
and the attractor equations \jji, the fully corrected entropy
formula \fdc\ is brought into the simple form
\eqn\fji{S_{BH}(q , p)
= {\cal F}(\phi , p ) - \phi^\L {\partial \over \partial
\phi^\L} {\cal F}(\phi, p). }
Here $q$ in the left-hand side and $\phi$ in the right-hand
side are related by \jji .
We recognize that \fji\ combined with \jji\ is the Legendre
transformation from $S_{BH}(q_\L, p^\L)$
to ${\cal F}(\phi^\L, p^\L)$ and that $\phi^\L$ are
chemical (electric) potentials for $q_\L$.

In order to interpret \fji\ we note that in the usual formulation
of a path integral, magnetic-type charges are associated to a
choice of vector bundle and are not summed over. On the other hand
electric-type charges are typically not fixed. Instead one chooses
a boundary condition on the electric potential, which gives a
weighted sum over electric charges. Hence the usual thermal path
integral should be interpreted as a microcanonical ensemble of
magnetic charges $p^\L$ and a canonical ensemble for electric
charges $q_\L$ with potentials $\phi^\L$. Such a ``mixed"
partition function can be written
\eqn\sxfl{Z_{BH}(\phi^\L,p^\L)=\sum_{q_\L}\Omega(p^\L,q_\L)e^{-
\phi^\L q_\L},}
where $\Omega(p^\L,q_\L)$ are integer black hole
degeneracies and $\ln \Omega(p^\L,q_\L)$
is the microcanonical entropy.

Hence we see that ${\cal F}(\phi^\L, p^\L)$ defined by \calf\
 is simply related to the mixed partition
function $Z_{BH}$ as
\eqn\dxzv{Z_{BH}(\phi^\L,p^\L)= \exp\left[{\cal F}(\phi^\L,
p^\L)\right],}
and that $S_{BH}(q_\L, p^\L)$ computed in \refs{\dwz, \dwa, \dwb,
\dwd} based on a
spacetime path integral picture is the mixed
canonical/microcanonical entropy given by the Legendre transform
${\cal F}$ as in \fji .
We note that, according to our proposal, the microcanonical entropy
$\ln \Omega(q_\L, p^\L)$ is $not$ equal the mixed entropy
$S_{BH}(q_L,p^\L)$ since the Legendre transformation is not
the inverse of the Laplace transformation \sxfl , except in
the limit of large electric charges $|q^\Lambda| \gg 1$,
where the latter can be approximated by the former in the
steepest descent method.

It is instructive to rewrite $Z_{BH}$ in terms of the complex
graviphoton charge $Q=Q_m+iQ_e$,\foot{This is a slight abuse of
terminology, as $Q$ is not precisely a charge. Rather it is a
combination of charges and potentials.} which is given as a
function of $\phi^\L$ and $p^\L$ by \eqn\ftyvm{Q^2=i {p^\L \over
4}(CF_\L-\bar C \bar F_{\L})+{\phi^\L \over 4\pi} (CF_\L +\bar C
\bar F_\L).} Choosing a gauge in which \eqn\dft{K=0,
~~~~~C=2Q,~~~~W^2={64 \over Q^2}} at the attractor point, we find
from \ftyv\ that \eqn\dxz{C=2Q,~~~~~X^\L={ p^\L+i\phi^\L/\pi \over
2 Q}, } and \eqn\rtfxp{\ln Z_{BH} = -4 \pi Q^2\Im \left[
\sum_hF_h\left( {p^\L+i{\phi^\L}/\pi\over 2 Q}\right)\left({8\over
Q}\right)^{2h}\right].}

\newsec{The topological string}

\subsec{Black hole and topological string partition functions}

The notion of topological string was introduced in
\lref\WittenIG{
E.~Witten,
``On the structure of the topological phase of two-dimensional gravity,''
Nucl.\ Phys.\ B {\bf 340}, 281 (1990).}
%%CITATION = NUPHA,B340,281;%%
\WittenIG .  Subsequently a connection
between them and superstring was discovered: It was shown
in \refs{\bcov,\naret}, that the superstring loop corrected $F$-terms
\dxvl\ can be computed as topological string amplitudes. The
purpose of this subsection is to translate the supergravity
notation of the previous section to the topological string
notation.

The second quantized partition function for the topological
string may be written
\eqn\ztop{Z_{\rm top}(t^A,g_{\rm top})
= \exp\left[ F_{\rm top} (t^A,g_{\rm top})\right],}
 where
\eqn\ftg{F_{\rm top}(t^A,g_{\rm top})=\sum_{h}g_{\rm top}^{2h-2}\
F_{{\rm
top}, h}(t^A),}
and $F_{{\rm top}, h}$ is the $h$-loop topological string
amplitude. The K\"ahler moduli are expressed in the flat
coordinates
\eqn\flatcoord{t^A = {X^A \over X^0} = \theta^A + ir^A,}
where $r^A$ are the K\"ahler classes of the Calabi-Yau $M$
and $\theta^A$ are periodic
$\theta^A \sim \theta^A + 1$.

We would like to determine relations between supergravity
quantities and topological string quantities.
Using the homogeneity property \efx\ and the
expansion \dxvl , the holomorphic
prepotential in supergravity can be expressed as
\eqn\prepexp{
\eqalign{ F(CX^\Lambda, 256) & = (CX^0)^2
F\left( {X^\L \over X^0},
           {256 \over (CX^0)^2}\right) \cr
& = \sum_{h=0}^\infty (CX^0)^{2-2h} f_h(t^A) , }}
where $f_h(t^A)$ is related to
$F_h(X^\L)$ in \dxvl\ as
\eqn\rescale{f_h(t^A) = 16^{2h}
F_h\left({X^\L\over X^0}\right) .}
This suggests an identification of the form $f_h(t^A)
\sim F_{{\rm top}, h}(t^A)$ and $g_{{\rm top}} \sim (CX^0)^{-1}$.
For later purposes, we need precise relations between
supergravity and topological string quantities including
numerical coefficients. These can be determined by studying
the limit of a large Calabi-Yau space.

In the supergravity notation, the genus $0$ and $1$ terms in
the large volume are given by \dwa\
\eqn\dpkl{\eqalign{
F\left(CX^\L,256
\right) &= C^2 D_{ABC}{X^AX^BX^C \over X^0}-{1 \over
6}c_{2A}{X^A \over X^0}+\cdots \cr
&=(CX^0)^2 D_{ABC}t^A t^B t^C-{1 \over 6}c_{2A}t^A+\cdots ,}}
where \eqn\uio{c_{2A}= \int_M c_2 \wedge \alpha_A ,} with $c_2$
being the second Chern class of $M$, and $C_{ABC}=-6D_{ABC}$ are
the four-cycle intersection numbers. These terms are normalized so
that the mixed entropy $S_{BH}$ is given by \fdc . On the other
hand, the topological string amplitude in this limit is given by
\refs{\bcovone, \bcov}\foot{In \bcov\ it was found that
consistency requires that an additional factor of $1/2$ be
multiplied to the one-loop amplitude $F_{{\rm top},1}$ defined in
\bcovone . The factor $1/2$ comes from the $Z_2$ automorphism of
the worldsheet torus.}
\eqn\onelooptop{
F_{\rm top}
  = - {(2\pi)^3 i \over g_{\rm top}^2} D_{ABC}t^A t^B t^C
- {\pi i\over 12} c_{2 A} t^A + \cdots,}
The normalization here is fixed by the holomorphic anomaly
equations in \bcov , which are nonlinear equations for
$F_{{\rm top},h}$.
The same normalization of $F_{\rm top}$
is also used in the M theory interpretation \gova\ and in the
quantum foam interpretaion \rov .

Comparing the one-loop terms in \dpkl\ and
\onelooptop , which are independent of $g_{\rm top}$,
we find
\eqn\tyi{F(CX^\Lambda, 256) = -{2i \over \pi}F_{\rm top}
(t^A, g_{\rm top}).}
Given this, we can compare the genus $0$ terms to find
\eqn\jiop{ g_{\rm top} = \pm {4\pi i \over CX^0}
.}
This implies
\eqn\toppartition{
\ln Z_{BH} = - \pi\ {\rm Im}\left[ F(CX^\Lambda, 256) \right]
= F_{\rm top} + \bar F_{\rm top}.}
and
\eqn\fxop{Z_{BH}(\phi^\Lambda, p^\L)
=|Z_{\rm top}(t^A, g_{{\rm top}})|^2,}
with
$$ t^A = {p^A + i \phi^A/\pi \over p^0 + i\phi^0/\pi},~~
             g_{{\rm top}} = \pm { 4 \pi i \over p^0 + i\phi^0/\pi}.
$$
Such a simple relation between $Z_{BH}$ and $Z_{\rm top}$ calls
for an explanation. The remainder of this paper is devoted
to physical interpretation of this relation.

\subsec{A simplified derivation?}

In this section we propose a simpler supergravity derivation of
the relation
\eqn\mre{Z_{BH}=|Z_{\rm top}|^2.}
With the exception of one assumption, which we cannot a priori
justify, the proposed derivation may clarify the origin of such a
relation. However we will not carefully work out the details and
we omit the factors of 2 and $\pi$ in this section.

One main ingredient in this derivation is the observation of
\soka\ that the ${\cal N}=2$ supergravity coupled to vector
multiplets can be written as
\eqn\supv{S=\int d^4x d^4 \theta \
({\rm supervolume\ form}) + h.c.=\int d^4x \sqrt{-g}R+...
}
where the supervolume form in the above depends non-trivially on
curvature of the fields.  This reproduces the ordinary action
after integrating over $d^4\theta$ and picking up the $\theta^4$
term in the supervolume. In the context of black holes the
boundary terms accompanying \supv\ give the
classical black hole entropy.

We now come to the derivation of \mre . As was observed in
\refs{\bcov ,\naret}\ topological string computes the terms
\eqn\topco{F=\sum_{h= 0}^\infty
\int d^4x d^4 \theta F_h(X)
(W^2)^g+ c.c.}
There are various terms one can get from the above action after
integrating over $d^4 \theta$.  Let us concentrate on one of the
terms which turns out to be the relevant one for us: Take the top
components of $X^\Lambda$ and $W^2$, and absorb the $d^4 \theta $
integral from the supervolume measure as in \supv .   We will
work in the gauge $X^0\sim 1$ and thus $C\sim 1/g_{\rm top}$.
As noted before in the near-horizon black hole geometry in this
gauge the top component
$W^2\sim 1/C^2\sim g_{\rm top}^2$ and the $X^\Lambda$ are fixed by the
attractor mechanism. We thus have the black hole free energy
$$\eqalign{
\ln Z_{BH} &=\sum_{h= 0}^\infty
g_{\rm top}^{2h} F_{{\rm top},h}
(X^\Lambda/X^0) \int d^4x d^4\theta+c.c.\cr
&=\sum_{g= 0}^\infty
(g_{\rm top})^{2h-2} F_{{\rm top},h}(X^\Lambda/X^0)+ c.c. \cr &=
2 \ {\rm Re}\ F_{\rm top},}$$
where we have used the fact that $\int d^4x d^4\theta
\sim 1/g_{\rm top}^2$.  Upon
exponentiation this leads to \mre .

Here we have shown that if we consider one absorption of
$\theta^4$ term in \topco\ upon $d^4\theta$ integral we get the
desired result.  That there be no other terms is not obvious.  For
example another way to absorb the $\theta$'s would have given the
familiar term $R^2 F^{2g-2}$ where $F$ is the graviphoton field.
However, as has been observed in \refs{\dwz, \dwa,\dwb}\ such terms do
not contribute in the black hole background.  It would be nice to
find a simple way to argue why these terms do not contribute and
that we are left with this simple absorption of the $\theta$
integrals.

\newsec{The holographic CFT dual}

The near-horizon geometry of a four-dimensional BPS black hole is
the Robinson-Bertotti universe AdS$_2 \times$S$^2$. On general
grounds one expects a holographically dual conformally-invariant
quantum mechanics CFT$_1$, but such AdS/CFT dualities are not yet
well-understand for the general case. In the special case $p^0=0$
a CFT$_2$ dual is known \msw , where the black hole geometry is expected
to arise upon considering a specific circle compactification
(i.e. a quotient of AdS$_3$ geometry
%\StromingerYG
\ref\StromingerYG{
A.~Strominger,
``AdS$_2$ quantum gravity and string theory,''
JHEP {\bf 9901}, 007 (1999)
{\tt hep-th/9809027}.
%%CITATION = HEP-TH 9809027;%%
}).
These cases are discussed in section 4.2. However to obtain
the most general allowed black hole charges we could consider
more general realization of the branes.  In the context
of type IIA superstring this can be achieved by considering
the gauge theory on D6 branes wrapped
on Calabi-Yau, which is a particular topolgoically
twisted gauge theory studied in \inov , which will be briefly
reviewed in section 4.1.  Since the main
interest of this paper focuses on index computations the details
of the CFT are not essential. The index computation can be done
for the QFT even before flowing to the IR.
 The basic idea is that the dual QFT should involve the
gauge theory living on the brane which leads to the black hole.
Suppose the D brane in question has $d+1$ dimensional worldvolume.
Then we write the QFT as a theory
 on $V^d \times S^1$ and the IR limit of this theory should lead
to a dual CFT$_1$ on $S^1$ (we view $R^1$ as the infinite radius
limit of $S^1$). In such cases we would be referring to a $(d+1)$
dimensional QFT as the dual theory and sometimes we loosely call
it the ``CFT dual''.

 For the purposes of our mixed ensemble
we would like to realize the theory on the brane
to be such that the electric charges can be induced as excitations
in the corresponding QFT.
Below we will present two examples of such a theory.

\subsec{Gauge theory on the $D6$ brane}

Consider type IIA theory compactified on the Calabi-Yau
$M$. We consider the
$D6$ and $D4$ branes wrapped over cycles of Calabi-Yau as
magnetic charged states and $D0$ and $D2$ branes as electrically charaged
states.  More specifically we wrap $N$ D6 branes on the Calabi-Yau
$M$ and we get a $U(N)$ gauge theory on $M\times S^1$.  This
gauge theory is a topologically twisted version of
the maximally supersymmetric Yang-Mills theory on Calabi-Yau
(see \inov\ for details).  Moreover the lower brane charges can be viewed
as gauge excitations on the $D6$ brane, as is well known.
Thus we end up with a maximally supersymmetric
gauge theory in 7 dimensions on $M\times S^1$.
Fixing the $D6$ and $D4$ brane charges as $p^A$, we will denote
this theory by $T(M,p^A)$.  Here $p^0=N$.

 As we
will discuss below we will be interested in the limit where
the radius of $S^1$, $\beta\rightarrow 0$.  In this limit
the gauge theory simplifies by T-duality on the circle
to a gauge thoery on $N$ $D5$ branes wrapping $M$.  Precisely
this theory was studied in \inov\ in finding a dual description
of the topological string.  We will discuss how this case is related
to a special case of our conjecture in section 5.

\subsec{Wrapped 5-brane dual}
For the other holographic dual we set one of
the magnetic charges $p^0=0$.  In this case the holographic duality can be
simply understood in one higher dimension.
Briefly, (see \msw\ for details ) it is a
compactification of the AdS$_3$/CFT$_2$ duality associated to the
black string obtained by wrapping fivebranes on a four-cycle in
the class $p^A$ of a Calabi-Yau compactification of M theory.
Membrane charge corresponds to turning on the self-dual three form
flux on the 2-cycle $q_A$ in the fivebrane. 0-brane charge is
obtained on further compactification to IIA by adding momentum to
the black string wrapping the M-theory circle.

 The resulting dual CFT, which
we will denote $T(M,p^A)$ can be briefly described as follows
\msw. The four-cycle is dual to an element $P$ of $H^2(M,Z)$,
where $M$ denotes the Calabi-Yau and we assume $P$ corresponds to
a very ample divisor. Then the moduli space $\CM_P$ has dimension
\eqn\mdim{d_P={1 \over 6}\int_M\bigl(2P^3+Pc_2(M)\bigr)-2.} The
theory $T(M,p^A)$ contains $d_P$ bosons with target space
$\CM_P$. There are also 3 more nonchiral bosons from translation
in $R^3$. In addition reduction of the self-dual three-form living
on the fivebrane gives chiral bosons. The number of such left-moving bosons is
\eqn\mdim{b_2^-(P)={1 \over 6}\int_M\bigl(4P^3+5Pc_2(M)\bigr)-1,}
while the number of right movers is  \eqn\mdim{b_2^+(P)={1 \over
6}\int_M\bigl(2P^3+Pc_2(M)\bigr)-1.} These massless scalars live
on the signature $(b_2^+, b_2^-)$ Narain lattice
$\Gamma_P=H^2(P,Z)$.  Membrane charges $q_A$ correspond to momenta
in the signature $(1,b_2(M)-1)$ sublattice $\Gamma_M=H^2(M,Z)$ of
$\Gamma_P$. Hence we can regard $q_A$ as an operator constructed
from the chiral scalars in $T(M,p^A)$, while 0-brane charge is
simply  $q_0=L_0$.

The theory
$T(M,p^A)$  has $(0,4)$ supersymmetry, and accordingly the number
of right moving fermions is \eqn\dxs{N^F_R={1 \over
3}\int_M\bigl(2P^3+Pc_2(M)\bigr).} Assuming $b_1(M)=0$ there are
no left moving fermions.

We note that a priori $T(M,p^A)$ depends on all the moduli of $M$
(as well as the integer magnetic charges $p^A$) through the
moduli-dependence of $\CM_P$. The electric
charges $(q_0,q_A)$ arise as momentum operators in the theory associated
the chiral scalars spanning the Narain lattice $\Gamma_P$.

\subsec{The supersymmetric index}
The holographic duals we have described above
both contain the center of mass degree of freedom.  In comparison
with the ${\rm AdS}^2\times {\rm S}^2$ geometry it is natural to mod out
the center of mass degree of freedom.  This gets rid
of an infinite volume factor as well as fermionic zero modes associated
with the center of mass degree of freedom.
 We then define the supersymmetric
index for this theory by

\eqn\dpar{Z_{CFT}(p^A, \phi^\Lambda; \beta, M)={\rm Tr
}'\bigl[ (-)^{F} \exp(-\beta H-\phi^\L q_\L)\bigr].}

In the wrapped fivebrane description $F=2J^3_R$.
Naively one might think that $Z_{CFT}$, being an index, is
invariant under all smooth deformations of $T$, in particular
variations of the complex structure and K\"ahler moduli of $M$, and
also independent of the inverse temperature $\beta$. In
fact this is $not$ obviously the case for the following reason.
Consider for example the contributions with $q_\L=0$.
For the wrapped 5-brane dual, the moduli space $\CM_P$ of the four cycle $P$ has points at which the
single four cycle degenerates into two four-cycles. These points
are junctions with Coulomb branches at which the four cycle can
fragment into two four-cycles which move apart in the $R^3$.
For the $U(N)$ gauge theory on $D6$ brane the same phenomenon happens
where there are points where $U(N)\rightarrow U(K)\times U(N-k)$.  Thus in either case
the holographic dual theory develops a new noncompact direction and a new continuum in the
spectrum.

It is well-known that the number of BPS states can
jump in such a case; for example, it was observed in the case
of ${\cal N}=2$ gauge theories in four dimensions in
\sw , and it is realized in the Calabi-Yau
context in \ref\lerche{S.~Kachru, A.~Klemm, W.~Lerche, P.~Mayr and C.~Vafa,
`Nonperturbative results on the point particle limit of ${\cal N}=2$
heterotic string
compactifications,''
Nucl.\ Phys.\ B {\bf 459}, 537 (1996),
{\tt hep-th/9508155}\semi
A.~Klemm, W.~Lerche, P.~Mayr, C.~Vafa and N.~P.~Warner,
``Self-dual strings and ${\cal N}=2$ supersymmetric field theory,''
Nucl.\ Phys.\ B {\bf 477}, 746 (1996),
{\tt hep-th/9604034}.
%%CITATION = HEP-TH 9604034
%%CITATION = HEP-TH 9508155;%%
}. When this happens via the
emergence of a noncompact direction, it is expected that
an index becomes a continuous function of background moduli.
This is shown explicitly in two-dimensional models \ref\nsi{S.~Cecotti,
P.~Fendley, K.~A.~Intriligator and C.~Vafa,
``A new supersymmetric index,''
Nucl.\ Phys.\ B {\bf 386}, 405 (1992),
{\tt hep-th/9204102}.
%%CITATION = HEP-TH 9204102;%%
},
where the index ${\rm Tr} (-1)^F F e^{-\beta H}$ defined
for Landau-Ginzburg theories  turns out to be a continuous function
of moduli as it receives anomalous contribution from multiparticle
states, even though the number
of BPS states jumps in this case.

This fragmentation phenomenon is generic \mms\ and occurs at many
values of the charges.\foot{A closely related
phenomenon of multiple basins of attraction
 was pointed out in
\lref\denef{
F.~Denef,
``Supergravity flows and D-brane stability,''
JHEP {\bf 0008}, 050 (2000), {\tt hep-th/0005049}.
%%CITATION = HEP-TH 0005049;%%
}
\lref\Mooreone{
G.~W.~Moore,
``Arithmetic and attractors,''
{\tt hep-th/9807087}.
%%CITATION = HEP-TH 9807087;%%
}
\lref\Mooretwo{
G.~W.~Moore,
``Attractors and arithmetic,''
{\tt hep-th/9807056}.
%%CITATION = HEP-TH 9807056;%%
}
\refs{\Mooreone,\Mooretwo,\denef}.}
We thus expect that $Z_{CFT}$ acquires anomalous
dependence on the moduli of $M$. We will discuss this issue
in more detail in secton 7.
The index \dpar\ can also depend on the inverse temperature $\beta$,
as we will discuss in the next section. We will argue that the mixed
partition function of black holes discussed in this paper is
identified with the $\beta \rightarrow 0$ limit of the index.

\newsec{Counting of BPS states and topological string}

The aim of this section is to consider some possible
checks for
the conjecture $Z_{BH}=|Z_{\rm top}|^2$, by using the microscopic
description of the black hole entropy.

 Given
that $Z_{BH}$ is expressed in terms of the topological string
partition function as $|Z_{\rm top}|^2$,
it is natural to expect that we are computing an index,
$i.e.$, the number of black hole microstates of a given
set of charges where we count each
BPS multiplet with a $\pm$ sign correlated with whether
the lowest component is a boson or a fermion. An obvious
candidate would be the index \dpar\ discussed in the last
section.

Due to the presence of noncompact directions,
the index \dpar\ may depend
on $\beta$. Thus, we should ask what is the natural value for $\beta$
from the perspective of the black hole entropy.  We propose
that for the four-dimensional black hole under study, we should
take the trace over all states of a given charge with some
electric potential, {\it regardless} of their energy.  In other
words we should take the limit $\beta \rightarrow 0$.
According to what we have found in this paper, we would then
identify
\eqn\smallb{
\lim_{\beta \rightarrow 0} Z(p,\phi; \beta, M)
=Z_{BH}=|Z_{\rm top}(g_{\rm top}, t; M )|^2,}
where
$$ g_{\rm top} \sim {1\over C X^0}, ~~t^A \sim {X^A\over X^0}, $$
with $X$'s are fixed by the attractor mechanism as
$$ X^A \sim p^A + i {\phi^A \over \pi}. $$

Recently it was shown that the topological string partition function
$Z_{\rm top}$ for a toric Calabi-Yau manifold
is expressed in the following form  \refs{\rov, \inov},
$$ Z_{\rm top} = \sum_{q} \Omega(q) \exp( - g_{\rm top} \ q_0 - t^A q_A),$$
where $\Omega(q)$ is an integer given by the number of
configuration of a classical statistical model. We will find that,
when $p_\L=0$, the limit of the black hole entropy $\lim_{\beta
\rightarrow 0} Z_{BH}(p, \phi; \beta, M)$ can also be brought into
this form via a chain of duality transformations of Type II string
theory. More precisely we argue that the inclusion of the
anti-holomorphic contributions to the gauge theory description of
\inov\ leads to $|Z_{\rm top}|^2$ instead of $Z_{\rm top}$. More precisely
we argue that the QFT dual of our black hole in the
$\beta\rightarrow 0$ and by a combination of T and S duality gets
mapped to the same QFT as discussed in \inov\ where the
transformed chemical potentials take the values given above.  Thus
the crystal melting ensemble is the S-dual of the mixed black hole
ensemble.

\subsec{A realization of the D-brane theory for Type IIA superstrings}

Consider Type IIA superstring compactified on a Calabi-Yau
threefold.  In order to formulate the partition function
\dpar\ in the superstring context, we take a Euclidean
time and compactify it on a circle of radius $\beta$ with
periodic (supersymmetry preserving) boundary conditions
for the fermions.  We consider a system of $D$6 and $D$4
branes carrying magnetic charges $p^0$ and $p^A$,
and regard $D$0 and $D$2 branes as electric excitations.
\lref\baulieu{
L.~Baulieu, H.~Kanno and I.~M.~Singer,
``Special quantum field theories in eight and other dimensions,''
Commun.\ Math.\ Phys.\  {\bf 194}, 149 (1998),
{\tt hep-th/9704167}\semi
%%CITATION = HEP-TH 9704167;%%
B.~S.~Acharya, M.~O'Loughlin and B.~Spence,
``Higher-dimensional analogues of Donaldson-Witten theory,''
Nucl.\ Phys.\ B {\bf 503}, 657 (1997),
{\tt hep-th/9705138}\semi
%%CITATION = HEP-TH 9705138;%%
M.~Blau and G.~Thompson,
``Euclidean SYM theories by time reduction and special holonomy  manifolds,''
Phys.\ Lett.\ B {\bf 415}, 242 (1997),
{\tt hep-th/9706225}.
%%CITATION = HEP-TH 9706225;%%
}
\lref\donaldson{S.~Donaldson and R.~Thomas, ``Gauge theory in
higher dimensions,'' in {\it The geometric universe;
science, geometry, and the work of Roger Penrose,}
S. Huggett et. al eds., Oxford Univ. Press. 1998.}
The theory on the branes is a gauge theory, which is topologically
twisted \ref\bsv{M.~Bershadsky, C.~Vafa and V.~Sadov,
``D-branes and topological field theories,''
Nucl.\ Phys.\ B {\bf 463}, 420 (1996),
{\tt hep-th/9511222}.
%%CITATION = HEP-TH 9511222;%%
} to give the right amount of supersymmetry. 
Gauge theories of this type have been discussed
in \refs{\baulieu,\donaldson}. We will consider the
case with $p^\L=0$ since this will simplify some of the analysis
below.  In this context we should view the gauge theory as
$U(N|N)$ where the lower brane charges are induced as excitations
of this gauge theory.

The index \dpar\ can be computed either as a trace over the QFT
Hilbert space  (with the center-of-mass degrees of freedom removed
in the case of \dpar ) or by periodically identifying the
imaginary time of Euclideanized AdS$_2$ with period $\beta$. From
the latter point of view, the index is given by a sum over BPS
states in AdS$_2$. If the type IIA string coupling $g_{\rm A}$ is
small, dominant contributions would come from $D0$ and $D2$ branes
carrying electric charges $q_0$ and $q_A$ in the form,
\eqn\impexp{{\rm exp}\left( - {\beta \over g_{\rm A}}q_0
 -{\beta \over g_{\rm A}}t^A q_A\right),}
where $t^A$ is the K\"ahler class measured in the type IIA
string frame.

Let us consider the limit $\beta \rightarrow 0$, where we expect
the identification \smallb\ works. In order for the
$D$0 and $D$2 branes to give finite contributions, we
also take $g_{\rm A} \rightarrow 0$ keeping the ratio
$\beta/g_{\rm A}\ll 1$.
Comparing this with the corresponding contribution
in \dpar\ from states with charges $q_0, q_A$, we can identify
the type IIA variables $g_{\rm A}, t^A$ with the chemical
potentials $\phi^0, \phi^A$ as
\eqn\adscft{ \phi^0 = {\beta\over g_{\rm A}},~~
\phi^A={\beta \over g_{\rm A}} t^A. }
The contribution of the $D$0 branes also suggests the identification
of the type IIA string coupling with the topological string
coupling as
\eqn\iiatop{ g_{\rm top} \sim {g_{\rm A} \over \beta}.}

Note that this gives an alternative derivation of
the attractor mechanism.
As we have shown in section 3, $X^0, X^A$ are
related to topological string parameters as
\eqn\xtop{
X^0 \sim {1 \over g_{\rm top}}, ~~ X^A \sim {t^A \over g_{\rm top}}.}
Assuming \iiatop ,
we arrive at the relation,
$$ \eqalign{ X^0 & \sim {1 \over g_{\rm top}}
\sim {\beta \over g_{\rm A}} = \phi^0, \cr
X^A & \sim {t^A \over g_{\rm top}}
 \sim {\beta \over g_{\rm A}} t^A = \phi^A .}$$
This reproduces the attractor values \chemical\ of the
moduli.

Note however, that in the limit of small $\beta$ it
is natural to do a $T$-duality.
Taking $T$ dual around the imaginary time direction
gives Type IIB theory on a circle of radius $1/\beta$
and with string coupling
$g_{\rm B} = g_{\rm A}/\beta$.   As already mentioned we are
interested in the limit $\beta/g_{\rm A} \ll 1$ which implies
that $g_{\rm B} \gg
1$.
Thus to get a better description of this type IIB theory we need to do
an S-duality.
After $S$ duality
$g_{\rm B} \rightarrow 1/g_{\rm B}$.  Note that under S-duality
$t^I/g_{\rm B}\rightarrow t^I.$
By doing this S-duality the amplitudes of topological
(A-model) string get related to amplitudes
of D(-1) and D1 brane instantons in the CY.
This is related to the fact that topological string
partition function has an expansion of the form
\eqn\ztop{Z_{\rm top}=\sum_{q} \Omega (q)
\exp(-g_{\rm top} q_0 - t^A q_A),}
where in the latter $q_0$ and $q_A$ can be interpreted as the D-instanton
and D1-brane contribution in the type IIB theory.  This result
for topological string was recently
derived in \refs{\rov,\inov},
for toric Calabi-Yau manifolds,
where it was shown that $\Omega(q)$ counts a number of configurations
in a classical statistical model describing melting
of a crystal.   More precisely $\Omega(q)$ is related
to the partition function of the theory on a $D5$ brane wrapping
the Calabi-Yau with fixed induced $D0$ and $D2$
brane charge--exactly the same theory which we have
discussed as a holographic dual to the black hole.
Moreover it was conjectured in \inov\ and in
\ref\pandet{D. Maulik, N. Nekrasov, A. Okounkov
and R. Pandharipande, ``Gromov-Witten theory and Donaldson-Thomas theory,''
{\tt math.AG/0312059}.}\ that this relation is true
for general compact Calabi-Yau as well.
We thus find that $\Omega(q)$ is related to the number
of BPS states of the black hole with $p=0$.
This relation between states
of the topologically twisted theory on the brane and
the topological string partition function
has been recently explained
in \nov\ by using S-duality of Type IIB
superstrings, which is T-dual to the S-duality we have used
here.\foot{See \ref\KapustinJM{
A.~Kapustin,
``Gauge theory, topological strings, and S-duality,''
{\tt hep-th/0404041}.
%%CITATION = HEP-TH 0404041;%%
} for discussion of a related issue from a different
point of view.}
% The specific case considered in \inov\ involved one $D$6 brane
%(leading to a $U(1)$ gauge theory on the Calabi-Yau),
%but it could also have been formulated without the D6 brane.
The relation \ztop\ is similar the relation we are anticipating
from our general analysis applied to this special case.  Namely
the right hand side of \ztop\ is similar to what we have called $Z_{BH}$
for this holographic dual.
However there are two important differences:
This equation relates
$Z_{\rm crystal}=Z_{\rm top}$ instead of $Z_{BH}=|Z_{\rm top}|^2$.  We believe
this may be due to the fact that in the analysis
of \inov\ one was specifically considering holomorphic
moduli dependence of the holographic
dual Yang-Mills theory on Calabi-Yau.  But there is a more
fundamental difference between the two ensembles:
The black hole ensemble and the  crystal melting ensemble
are S-dual of one another ($i.e.$, we have a strong/weak duality on
the topological string coupling constant) and so {\it there is no
one to one map from a melting crystal configuration to that of
a black hole}.  Nevertheless, it would be important to study this relation
further.

Note that the connection between topopological strings
and topologically twisted Yang-Mills theory we have
proposed is analogous to that of $N=2$
topological Yang-Mills theory in 2 dimensions.  In that case
it is known that $N=2$ topological Yang-Mills theory is equivalent
to ordinary non-supersymmetric Yang-Mills theory \ref\wittentym{
E.~Witten,
``Two-dimensional gauge theories revisited,''
J.\ Geom.\ Phys.\  {\bf 9}, 303 (1992),
{\tt hep-th/9204083}.
%%CITATION = HEP-TH 9204083;%%
}.  Moreover the ordinary
Yang-Mills theory has been studied in the limit of large rank
\lref\gross{
D.~J.~Gross,
``Two-dimensional QCD as a string theory,''
Nucl.\ Phys.\ B {\bf 400}, 161 (1993),
{\tt hep-th/9212149}.
%%CITATION = HEP-TH 9212149;%%
}\lref\gta{
D.~J.~Gross and W.~I.~Taylor,
``Two-dimensional QCD is a string theory,''
Nucl.\ Phys.\ B {\bf 400}, 181 (1993),
{\tt hep-th/9301068};
%%CITATION = HEP-TH 9301068;%%
``Twists and Wilson loops in the string theory of two-dimensional QCD,''
Nucl.\ Phys.\ B {\bf 403}, 395 (1993),
{\tt hep-th/9303046}.
%%CITATION = HEP-TH 9303046;%%
}
\lref\Cordes{
S.~Cordes, G.~W.~Moore and S.~Ramgoolam,
``Large $N$ 2-D Yang-Mills theory and topological string theory,''
Commun.\ Math.\ Phys.\  {\bf 185}, 543 (1997),
{\tt hep-th/9402107}.
%%CITATION = HEP-TH 9402107;%%
}
\refs{\gross,\gta,\Cordes},
which is analogous to our semiclassical expansion of black holes,
and it has been seen that it can be written
as a square of a topological theory, which maps worldsheet Riemann
surfaces holomorphically to the target 2d space where the
Yang-Mills lives on.  They find for the 2D Yang-Mills case
$$Z_{YM}=|Z_{\rm top}|^2,$$
very much in the spirit of what we are predicting
for the higher dimensional topologically twisted version
of YM on CY.

Let us now turn to the other limit $\beta \rightarrow \infty$.
In this case, only ground states can contribute to \dpar .
In order for $D$0 and $D$2 branes contribute appreciably,
as is evident from \dpar , we need to take $g_{\rm A}
\rightarrow \infty$ so that $g_{\rm A}/\beta$ stays finite.
Thus, it is natural that the $\lim_{\beta \rightarrow\infty}Z$
is related to a 5d black
hole computation in M theory on the Calabi-Yau space times
a circle.   This
is in fact the duality explored in \gova\
where M2 brane degeneracies was connected to topological A-model
strings.  More precisely in this five-dimensional context one identifies
the $t^A$ chemical potential, as the chemical potential
for the $M2$ brane, and the $g_{\rm top}$ with the chemical potential
for the $J_3$ generator in $SU(2)_L$ subgroup of the $SU(2)_L\times
SU(2)_R=SO(4)$ spatial rotation group in five dimensions.

It is natural also to ask what does the $\beta $ dependence of $Z$
mean for the black hole entropy.  It is morally a
black hole in the background of ``supersymmetric temperature $1\over \beta$''
(due to periodic fermionic boundary conditions).

\newsec{$Z_{\rm top}$ as a quantum wave function}
 The relation $Z_{BH}=|Z_{\rm top}|^2$ suggests that
 $Z_{\rm top}$ should be interpreted as a quantum state
 and
$Z_{BH}$ the corresponding probability amplitude. Roughly
speaking, $Z_{\rm top}$ is the quantum amplitude for finding a black
hole with charges $(p^\L,q_\L)=(\Re\ CX^\L, \Re\ CF_\L)$.

This fits nicely with the discussion in \wittenqb , where it was
observed that the holomorphic anomaly of \bcov\ can be
interpreted in a simple way if one identifies
$Z_{\rm top}$ as a state in an ``auxiliary"
Hilbert space arising from geometric quantization of the moduli
space $\CM_M$ of the Calabi-Yau $M$ as a phase space:
In the B-model context we define
$$CX^\L=\int_{A_\L} \Omega ,\qquad C F_\L=\int_{B^\L} \Omega$$
where $A_\L, B^\Sigma $  are symplectic conjugate 3-cycles and
assign a commutation relation which in our normalizations is given
by
\eqn\comut{[C X^\L,C F_\Sigma]={2i \over \pi} \delta^\L_\Sigma,}
or equivalently
$$ [X^\L, F_\Sigma]={2i \over \pi C^2} \delta^\L_\Sigma. $$
The topological string amplitudes depend only on half of these
variables, which we take to by $X^\L$. The choice of the
background Calabi-Yau $M_0$ which one expands about changes what
one means by $\Omega$ and thus how one writes the $Z_{\rm top}(X^\L)$.
This dependence on $M_0$ was proposed in \wittenqb\ to be
equivalent to the holomorphic anomaly equation of \bcov .  This
relation has been further elucidated recently in \mvv .  Note that
$Z_{\rm top}$ not only depends on the background (whose black hole
interpretation we will discuss in the next section) but it also
depends on the choice of a symplectic basis of 3-cycles ($i.e.$ 
the separation of 3-cycles into the A and B cycles).  In this
section we would like to explore, for a fixed background $M_0$ the
dependence of $Z_{\rm top}$ on the choice of the symplectic basis.

Note that \comut\ is compatible with the semiclassical statement
that $F_\Sigma =\partial_\Sigma F_0(X^\L)$.  To see this recall
that in our normalization we have
$$\eqalign{ e^{F_{\rm top}}& ={\rm exp}
\left[ {\pi i \over 2} C^2 F\left(X, {256 \over C^2}\right) \right] \cr
&={\rm exp} \left[ {\pi i \over 2} C^2 F_0(X) + \cdots \right],}$$
Note that $1/C$ plays the role of the topological string coupling
constant $g_{\rm top}$. Thus, representing
$$F_\Sigma =-i{2\over \pi C^2} {\partial \over \partial X^\Sigma }$$
acting on the topological string wave function and taking the 
semi-classical limit 
$C\rightarrow \infty$ ($i.e.$, $g_{\rm top} \rightarrow 0$), we recover the
classical relation $F_\Sigma =\partial_\Sigma F_0(X^\L)$.

In our context we also have to deal with the anti-topological sector.
The same analysis goes through.  To keeps the sign conventions
compatible with our previous discussion note that
$$e^{{\overline F}_{\rm top}}={\rm exp}
\left[-i{\pi \over 2} \bar C^2 {\overline F_0}({\overline X^\L})+
\cdots \right]$$
which is compatible with the commutation relation
\eqn\comut{[{\overline X^\L},{\overline F_\Sigma }]=- {2i \over
\pi \bar C^2} \delta^\L_\Sigma}
Moreover the topological and anti-topological amplitudes are independent
of one another, which implies that all the barred variables commute
with all the unbarred variables.

How are these commutation relations compatible with our black hole
interpretation?  To see this recall 
$$\eqalign{& p^\L={\rm Re}\left[ C X^\L\right]\cr
& q_\L={\rm Re} \left[ C F_\L\right] .}$$
Using the above commutation relations we learn that, as expected
the electric and magnetic charges do commute:
$$[p^\L,q_\Sigma ]=0.$$
This is expected because we can measure both at the same time.
However, we can learn something new from these commutation
relations: So far we have talked about electric chemical
potentials $\phi^\L$. We could also define magnetic chemical
potentials $\chi_\L$.  Note that because of their definition one
expects these to be conjugate to the corresponding charges (this
is true if we replace sum over charges by integrals which is true
to all order in string expansion)
$$[\phi^\L,q_\Sigma ]=\delta^\L_\Sigma  \qquad [\chi_\Sigma , p^\L]=\delta_\Sigma^\L$$
$$[\phi^\L,p^\Sigma ]=0\qquad [\chi_\Sigma , q_\L]=0$$
For these to be compatible with the above commutation relation
we are led to the unique choice
$$\eqalign{& C X^\L = p^\L + {i \over \pi} \phi^\L \cr
& C F_\L = q_L + {i \over \pi} \chi_\L, }$$
which is exactly the relations we had obtained before. We thus see
that the structure of topological string anticipates the
identification of chemical potential with the imaginary parts of
the periods.

Now we come to another issue which is the dependence of $Z_{\rm top}$
on the choice of the symplectic basis. The fact that $Z_{\rm top}$
does depend non-trivially on a choice of symplectic basis is
perfectly compatible with the black hole interpretation of it:
The choice of the mixed ensemble also depends on which charges we
choose as electric and which ones we choose as magnetic.  However
topological string amplitudes tells us how the wave function
$Z_{\rm top}$ changes as we change this basis.  For example if we
switch the role of $A$ and $B$ cycles, $i.e.$ exchange electric and
magnetic charges, the topological string partition function goes
to its Fourier transform. What does this mean for our black hole
interpretation?

To gain insight into this let us turn off the chemical potentials.
So we have $X^\L\sim p^\L, F_\Sigma \sim q_\Sigma  $.  Let us
denote
$$Z_{\rm top}(X^\L)=Z_{\rm top}(p^\L)=\tilde\psi(p)$$
$$Z_{\rm top}(F_\Sigma )=Z_{\rm top}(q_\Sigma )=\psi(q)$$
The fact that exchanging A and B cycles conjugates the variables
implies that $\psi(q)$ and $\tilde\psi(p)$ are Fourier transform
of one another.  In other words
$$\psi(q)=\int dp \ \tilde\psi(p)\ {\rm exp}(2\pi i p^\L q_\L)$$
(note that this expression is well defined in the perturbative topological
string expansion where $q,p$ are not fixed to be integer).

Let us relate these to black hole partition function.  When we set
chemical potentials to zero we have
$$Z_{BH}=\sum_{q} \Omega(p,q)=|\tilde\psi(p)|^2$$
$$Z_{BH}=\sum_{p} \Omega(p,q)=|\psi(q)|^2$$

We can invert this relation as well.  Namely, if we
are given the topologial string amplitude $\psi(q)$ we could
ask how does this lead to $\Omega (p,q)$?    This can be done
by inverting the relation $Z_{BH}=|Z_{\rm top}|^2$ and we find (ignoring
for simplicity the discretization of charge, which is fine to all
orders in perturbative string expansion)
$$\Omega (p,q)=\int d \chi \ {\rm exp}(-p \chi)\
\left|\psi\left(q+i {\chi \over \pi}\right)\right|^2$$
The fact that the topological string amplitude transforms
as a wavefunction, $i.e.$ that $\phi(p)$ is Fourier transform of
$\psi(q)$ implies that even if we had chosen a different polarization
we would have gotten the same answer, in other words we can also write
$$\Omega (p,q)=\int d \phi \ {\rm exp}(-q \phi)\
\left|\tilde\psi\left(p+{i\over \pi} \phi
\right)\right|^2$$
Note that this relation between $\Omega $ and $\psi$ or $\phi$ is a
bit formal, and to makes sense of it we have to choose a suitable contour.
In fact this is the reason why $\Omega$ does not a
priori have to be
positive.
It is natural to choose the contour so it looks like a Fourier transform
and we have
$$\Omega(p,q)=\int d \chi\  {\rm exp}(-ip \chi)\
\psi\left(q+ {\chi \over \pi}\right)
\psi^*\left(q-{\chi \over \pi}\right)$$
Note that this relation implies that $\Omega$ is real, but
it is not necessarily positive.
We recognize $\Omega$ as the {\it Wigner function} associated
to the wave function $\psi$.  This is familiar in the reformulation
of quantum mechanics as a deformation quantization (see \ref\wig{
C.~K.~Zachos,
``Deformation quantization: Quantum mechanics lives and works in
phase-space,''
Int.\ J.\ Mod.\ Phys.\ A {\bf 17}, 297 (2002),
{\tt hep-th/0110114}.
%%CITATION = HEP-TH 0110114;%%
%J. Hancock, M. A. Walton, B. Wynder, ``Quantum mechanics another way,''
%{\tt physics/0405029}
}\ for a review
and further references).  In the context
of this reformulation of quantum mechanics the fact that Wigner's
function is a `quasi-probability distribution' refers to the fact that
$\Omega$ need not be positive.  For us this is also the case because
$\Omega$ is an index of the black hole.  Even though the lack of positivity
was a difficult thing to interpret in the context of Wigner's reformulation
of quantum mechanics, it is perfectly natural for us in the context
of the relation between the black hole and topological strings:
The black hole entropy is an {\it indexed} object.

These formulae are very suggestive and in need of a deeper
physical explanation:  Is black hole teaching us about
a reformulation of quantum mechanics?  Are the black hole
information puzzles
resolved in this new reformulation?

\newsec{Further Issues}

In most of the preceding sections we have ignored important
subtleties related to background dependence. These subtleties
arise in all three pictures. In the supergravity picture we have
black hole fragmentation, in the dual CFT picture we have multiple
Coulomb branches and in the topological string picture we have the
holomorphic anomaly. Although we have not understood these issues
in detail, we believe that they are different manifestations of
the same thing, and that a relation of the form
$Z_{CFT}=Z_{BH}=|Z_{\rm top}|^2$ will survive these corrections. This
section contains some preliminary discussion of these issues.

\subsec{Fragmentation and jumping lines} Usually it is assumed
that the number of BPS states of a single black hole depends only
on the charges of the black hole. However this is not quite true
for the case of 4d BPS black holes obtained from compactification
of Type II superstrings on Calabi-Yau threefolds. There are two
main issues. The first one is that it is known that the number of
BPS states preserving half of the ${\cal N}=2$ supersymmetries
does not depend just on the charges; it was already discovered in
\sw\ that the number of such states can jump as we change the
moduli of the scalars in the vector multiplet. This is likely
related to the black hole fragmentation discussed in \mms\ as well
as the phenomenon of multiple basins of attraction 
\refs{\Mooreone,\Mooretwo,\denef}. We
thus expect $Z_{BH}$ not only to depend on $X$, which captures
the magnetic charge and the electric potentials, but also on the
``background'', $i.e.$ the asymptotic value of the moduli in the
supergravity solution. The asymptotic vector multiplet background
is denoted by $t_V$. (Possible hypermultiplet dependence is also
discussed below.)

Next we ask whether or not $t_V$ dependence appears in the
supersymmetric index, \eqn\dfsx{{\rm Tr }'\bigl[
(-)^{F} e^{-\phi^\L q_\L}\bigr].} In doing so we
must specify whether the trace is over multi-black hole or single
black hole states. The spectrum of single black hole states
changes erratically as the asymptotic moduli cross basins of
attraction or jumping lines. Hence the trace will be a
discontinuous function of the background moduli. The other option
is to include multi-black-hole states in the trace. However in
that case the spectrum will in general have many continuous
parameters arising from the non-compact $R^3$ positions of the
black holes (which are nevertheless expected to yield finite
answers due to the index nature of the computation).
The index can still be defined in this context, but
it will depend on the asymptotic values of the moduli. Hence
either way we have background dependence. The best we can do is to
define the multi-black hole index which has a smooth background
dependence: \eqn\rzb{Z_{BH}=Z_{BH}(p^\L,\phi^\L;t_V).}

\subsec{The holomorphic anomaly} Now we come to the discussion of
similar issues for the topological strings.  It is known that the
perturbative topological string partition function depends on the
background they are expanded about.  This is encoded in the
holomorphic anomaly of topological strings \refs{\bcov ,\bcovone}.
Thus, at least perturbatively we know that
\eqn\rzbb{Z^{\rm pert.}_{\rm top} = Z^{\rm pert.}_{\rm top}(X^\Lambda;t_V).}
This fits with the expectation on $Z_{BH}$ in the above.
An important check would be to show that the $t_V$ dependence
in \rzb\ agrees with that of \rzbb .

\subsec{Coulomb branches} In the dual CFT picture we have the
graded partition function $Z_{CFT}$ of fivebranes with fluxes
wrapping Calabi-Yau four-cycles, or D6 branes wrapping the Calabi-Yau.
At certain points of their moduli space, the fivebranes/D6 branes can
fragment and move apart in $R^3$.  This again leads to dependence
on $t_V$.

\subsec{Hypermultiplets}

 The dependence on background vector multiplets is however not the end of the story; it is
also known that the number of BPS states do change even as we
change the hypermultiplet moduli. To give a simple example,
consider Type IIA strings compactified on the quintic threefold.
Suppose we wish to know the number of BPS $D$2 branes wrapping the
basic cycle of quintic once. It is known that the number of such
D2 branes does depend on the complex structure of the quintic. The
complex structure in this case is part of the hypermultiplet
moduli.  Thus we expect that the number of BPS states to also
depend on the background hypermultiplet moduli, which we denote by
$t_H$.  In other words it is already known that we should expect
the relevant black hole partition function to be a function of
$t_H$. However it is not clear whether or not this $t_H$
dependence persists when we consider the index.  Naively
one may think it disappears.

Some evidence that $t_H$ dependence may indeed persist is that
there exist natural non-perturbative completions of the
topological string which depend on $t_H$.  This is for example
natural in the twistor application of topological strings \wittwi\
where the D-instantons of the topological string play a role.  As
noted in \refs{\nv ,\nov} these D-branes couple to $t_H$, and thus
we expect that the full non-perturbative topological string to
also depend on it, in other words one might expect
non-perturbatively
$$Z_{\rm top}(X^\Lambda; t_V,t_H)$$

If so it would be natural to expect that our relation continues to
hold even after we take into account these subtleties on both
sides, taking the form
\eqn\funeq{Z_{BH}(X^\Lambda;t_V,t_H)=|Z_{\rm top}(X^\Lambda; t_V,t_H)|^2}
Consistency of this relation with $t_H$-independence of
perturbative topological string would require
that the dependence of the black hole entropy on hypermultiplets
is not present to all orders in the $(1/Q)$ expansion, and only
shows up in non-perturbative terms of the form $\exp(-Q)$.

It is also conceivable
that there are other completions of topological strings
which do not depend on $t_H$.  This would in particular
require setting the D-brane contributions to zero by hand,
as they clearly do couple to $t_H$.  If there were such
alternative non-perturbative definitions we could imagine
having
$$Z^{index}_{BH}(X^\Lambda;t_V)=|Z_{\rm top}(X^\Lambda,t_V)|^2.$$
 The question is analogous to asking ``can we define
alternative definitions of topological Type II superstrings in
which D-branes decouple?'' One might think this is unlikely and
that the $t_H$ dependence of $Z_{BH}$ should be there.  As is
clear we have just opened up a direction of investigation.  Much
more work is needed for a deeper understanding of the beautiful
link between topological strings and black holes.

\bigskip
\bigskip

\centerline{\bf Acknowledgements}

We would like to thank M. Aganagic, F. Denef, R. Dijkgraaf, M.
Douglas, G. Moore, A. Neitzke, M. Rocek, A. van Proeyen and E.
Witten for valuable discussions. The work of HO was supported in
part by DOE grant DE-FG03-92-ER40701, that of AS was supported in
part by DOE grant DE-FG02-96ER and that of CV was supported in
part by NSF grants PHY-0244821 and DMS-0244464.

\listrefs

\bye